\documentclass[aps,pra,twocolumn,amssymb,amsmath,%
    superscriptaddress,showpacs,showkeys]{revtex4}

\newcommand{\ket}[1]{|{#1}\rangle}
\newcommand{\bra}[1]{\langle{#1}|}
\newcommand{\braket}[2]{\langle{#1}|{#2}\rangle}
\newcommand{\HH}{{\mathcal H}}
\newcommand{\EE}{{\mathcal E}}
\newcommand{\id}{\openone}
\newcommand{\eref}[1]{Eq.~(\ref{#1})}
\newcommand{\sref}[1]{Sec.~\ref{#1}}
\newcommand{\Sref}[1]{Sec.~\ref{#1}}

\begin{document}
\title{Continuous variable remote state preparation}
\author{Z. Kurucz}
\affiliation{Department of Nonlinear and Quantum Optics, Research
  Institute for Solid State Physics and Optics, Hungarian Academy of
  Sciences, PO Box 49 H-1525 Budapest, Hungary}
\author{P. Adam}
\affiliation{Department of Nonlinear and Quantum Optics, Research
  Institute for Solid State Physics and Optics, Hungarian Academy of
  Sciences, PO Box 49 H-1525 Budapest, Hungary}
\affiliation{Research Group for Nonlinear Optics, Hungarian Academy of
  Sciences, and Institute of Physics, University of P\'ecs,
  Ifj\'us\'ag \'ut 6. H-7624 P\'ecs, Hungary}
\author{Z. Kis}
\affiliation{Department of Nonlinear and Quantum Optics, Research
  Institute for Solid State Physics and Optics, Hungarian Academy of
  Sciences, PO Box 49 H-1525 Budapest, Hungary}
\author{J. Janszky}
\affiliation{Department of Nonlinear and Quantum Optics, Research
  Institute for Solid State Physics and Optics, Hungarian Academy of
  Sciences, PO Box 49 H-1525 Budapest, Hungary}
\affiliation{Research Group for Nonlinear Optics, Hungarian Academy of
  Sciences, and Institute of Physics, University of P\'ecs,
  Ifj\'us\'ag \'ut 6. H-7624 P\'ecs, Hungary}

\begin{abstract}
  We extend exact deterministic remote state preparation (RSP) with
  minimal classical communication to quantum systems of continuous
  variables.  We show that, in principle, it is possible to remotely
  prepare states of an ensemble that is parameterized by infinitely
  many real numbers, i.e., by a real function, while the classical
  communication cost is one real number only.  We demonstrate
  continuous variable RSP in three examples using (i) quadrature
  measurement and phase space displacement operations, (ii)
  measurement of the optical phase and unitaries shifting the same,
  and (iii) photon counting and photon number shift.
\end{abstract}
\pacs{03.67.Hk}%
\keywords{quantum computation, quantum communication, deterministic
  oblivious continuous variable remote state preparation,
  teleportation}%
  
\maketitle

\section{Introduction}
\label{sec:intro}

In remote state preparation (RSP), quantum
communication is combined with quantum state engineering.  The sender
(Alice), having a classical description of the state, prepares a
physical instance of its at the receiver's distant location (at Bob's
laboratory) using previously shared entanglement and classical
communication as resources.  A plausible protocol implementing this
task is the one in which Alice prepares the state locally in her lab
and teleports it to Bob~\cite{prl70p1895}.  This protocol consumes 1
ebit of entanglement and communicates 2 bits of classical information
per each qubit prepared.  However, it may be preferable to avoid
teleportation in certain situations, especially, if it is difficult to
implement the Bell-type measurements essential for the protocol.

Recently, several non-teleportation-based RSP protocols have been
developed \cite{tit51p56,pra62e012313,
  prl87e197901, prl90e127905, pra67e052302, pra68e062319,
  pra63e014302, pra65e022316, pra69e022310, job7p135}.  
They exploit the fact that Alice has complete classical
knowledge of the state to be prepared.  This gives the possibility to
trade off the resources.  For example, an alternative for Alice is to
tell everything to Bob so that he can prepare the state locally in his
lab.  This method requires no prior entanglement but the transfer of
two real numbers, that is, infinitely many bits of classical
information.  In the other limit posed by causality, it is possible to
communicate an arbitrary qubit using one classical bit only.  The
asymptotic protocol $\Pi$ of Ref.~\cite{tit51p56} achieves this rate
utilizing 1 ebit per transmitted qubit.  Besides these two extreme
cases, trade-off among resources has been widely investigated in the
literature \cite{pra62e012313, prl87e197901, prl90e127905,
  pra67e052302, pra68e062319}.

A conceptually different kind of RSP was introduced in
Ref.~\cite{pra63e014302}.  This achieves a trade-off from another
point of view: resources can also be cut down if the input is
restricted to an ensemble of special pure qubit states.  For example,
equatorial or polar great circles on the Bloch sphere can be prepared
remotely with probability 1 using one ebit of entanglement and 1 bit
of classical communication.  The protocol does not suffer from the
difficulties associated with the Bell measurement of quantum
teleportation because Alice performs a simple onepartite measurement
on her half of the entangled pair, projecting its spin along some
well-defined direction.  Furthermore, instead of the four unitaries of
the teleportation protocol, Bob applies either the identity
or~$\sigma_z$.

This kind of RSP has recently been generalized to the $N$ dimensional
case \cite{pra65e022316, pra69e022310, job7p135}.  The number of
possible outcomes yielded by the projective measurement is $N$ (in
contrast with $N^2$ in teleportation) and the unitaries Bob need to
implement form a commutative subgroup of the Weyl group of
teleportation, namely, it is the cyclic group $\mathbb Z_N$.
The ensemble of states that can be
prepared is an $N-1$ dimensional real manifold.  Since an arbitrary
state is given by twice as many parameters [$2(N-1)$ real numbers],
this kind of RSP realizes remote preparation of
``$\frac12$~qubit'' using 1 ebit of entanglement and 1 bit of
classical communication as resources.  It is also called
\emph{minimum} RSP.

The protocols mentioned so far deal with finite dimensional quantum
systems.  Nevertheless, light pulses used for quantum communication are
essentially described by continuous variables (CV), and CV quantum
information processing provides an interesting alternative to the
traditional qubit-based approach \cite{prl82p1784, pra64e012310,
  pra65e042304, prl88e097904, prl90e117901}.

In CV teleportation \cite{pra49p1473, prl80p869, sci282p706}, 
the nonlocal resource shared by Alice and Bob is
the Einstein--Podolsky--Rosen state \cite{pr47p777} with perfect
correlation in both position and momentum.  In a quantum optical
context, such a correlation can be approximated with a highly squeezed
two-mode state of the electromagnetic (EM) field with quadrature
amplitudes of the field playing the role of position and momentum.
Bell measurement at Alice's site is changed to the simultaneous
measurement of the center-of-mass position $\hat X \equiv \hat X_1 +
\hat X_{\text{in}}$ and relative momentum $\hat P \equiv \hat P_1 -
\hat P_{\text{in}}$ of Alice's half of the EPR pair and the input
particle.  Finally, the unitary operations Bob has to implement are
the phase space displacement operators $\hat D(\alpha)$ with
$\alpha=X+iP$ obtained from the measurement result.  They form the
well-known Heisenberg--Weyl group.

Though CV systems play an important role in quantum information
\cite{pra49p1473, prl80p869, sci282p706, prl82p1784, pra64e012310,
  pra65e042304, prl88e097904, prl90e117901}, a comprehensive study of
continuous variable remote state preparation (CVRSP) is still missing.
There are many quantum state engineering schemes based on conditional
measurements on one of two entangled light beams.  However, they are
not RSP, since they are probabilistic.  And it is an essential feature
of RSP that Bob's action makes it deterministic.  The scheme of
Ref.~\cite{job5pS360} can remotely prepare a squeezed state using
homodyne detection and it can be made deterministic by Bob applying a
phase space displacement operation conditioned on Alice measurement
result.  Still, only a restricted ensemble of states can be prepared,
since Alice has control over one (complex) parameter only.

In the present paper, we will show that by allowing Alice to perform
other kinds of measurements, we can enlarge the ensemble of preparable
states while letting Bob use the same set of unitaries and keeping the
classical communication cost.  The ensemble will be parameterized by
not only one but continuously many real numbers, that is, by real
functions.  We will present two other CVRSP protocols involving photon
counting and phase measurements that are based on the minimum RSP
scheme of \cite{pra65e022316, pra69e022310} by taking the infinite
limit in the dimensionality of the state spaces.

The paper is organized as follows.  In \sref{sec:finitersp}, exact
deterministic oblivious remote state preparation of finite dimensional
systems with minimal classical communication is introduced.  We
characterize the non-maximally entangled resource by antilinear
operators.  Then we give the ensemble in the $N$ dimensional case that
can be remotely prepared if Bob applies a special set of unitary
operations, namely, the group $\mathbb Z_N$ of integer addition modulo
$N$.  \Sref{sec:cvrsp} presents our main idea, we extend our analysis
to CVRSP, a topic in quantum information processing that has not yet
been thoroughly investigated.  We present three example protocols, one
based on quadrature amplitude measurement, the others involving
measurement of the optical phase and the number operator.
\Sref{sec:conc} summarizes our results.

\section{Minimum RSP in finite dimension}
\label{sec:finitersp}

The RSP protocol introduced in Ref.~\cite{pra63e014302} consists of
two steps: (i) Alice performs a projective measurement on her half of
the shared entangled pair according to the \emph{target} state that
she wants to prepare remotely at Bob's site, and communicates its
result to Bob; (ii) Bob applies a unitary transformation on his half
according to Alice's message to restore the target state.  Initially,
they share a pure (but not necessarily maximally) entangled state
\begin{equation}
  \label{eq:Psi}
  \ket\Psi_{AB} = \sum_{k=0}^{N-1} \alpha_k \ket{k}_A \ket{k}_B
\end{equation}
where $\alpha_k$ and $\ket{k}$ are the Schmidt coefficients and the
corresponding Schmidt vectors.  It is very intuitive to view the
entangled nature of the two subsystems through the antilinear
(conjugate linear) operator
\begin{equation}
  \label{eq:defRSO}
  R \colon \HH_A \to \HH_B, \quad 
  R\ket\phi_A \equiv {}_A \braket\phi\Psi _{AB},
\end{equation}
where the partial scalar product is antilinear in its first argument.
To illuminate its physical motivation, suppose that Alice finds her
part in a state $\ket\phi_A$ after a projective (von Neumann)
measurement.  Because of entanglement, Bob's state, conditioned on
this measurement outcome, reads
\begin{equation}
  \label{eq:psib}
  \ket\psi_B = \frac1{\sqrt p} {}_A \braket\phi\Psi _{AB} 
  = \frac1{\sqrt p} R\ket\phi_A,
\end{equation}
where the normalizing factor is obtained from the probability $p$
of this measurement event
\begin{equation}
  \label{eq:prob}
  p = \| {}_A \braket\phi\Psi _{AB} \|^2 = \|R\ket\phi_A\|^2.
\end{equation}
The entangled state $\ket\Psi_{AB}$ is completely given by the
antilinear operator (\ref{eq:defRSO}) that is mapping a possible
measurement eigenstate in system A to the corresponding state of
system B after the measurement.  This isomorphism between pure
entangled states and antilinear operators, and its application to
quantum teleportation has been thoroughly investigated in Refs.\
\cite{lnp539p93, fdp49p1019, job5pS627}.  Note that the polar
decomposition of the antilinear operator (\ref{eq:defRSO}) is
\begin{equation}
  \label{eq:antipolar}
  R=\sqrt{\rho_B}J,  
\end{equation}
where $\rho_B$ is the reduced density operator of the entangled
resource and $J \colon \HH_A \to \HH_B$ is the antiunitary isomorphism
that maps Schmidt vectors of system A to those of system B.

Now, \emph{any} target state $\ket\psi_B$ can be prepared remotely in
a probabilistic way if there is a state $\ket\phi_A$ in $\HH_A$ that
satisfies (\ref{eq:psib}) and Alice manages to project her system onto
this state.  However, the ensemble $\EE$ of target states that can be
prepared in a deterministic exact way is rather restricted due to the
fact that Bob can apply only a given set of unitary operations $U_j$
on his system in case Alice's measurement fails to result the state
$\ket\phi_A$.  It is easy to see \cite{pra65e022316, pra69e022310,
  job7p135} that RSP is possible in the special case when
\begin{enumerate}
\item[(i)] Bob's unitaries commute with the partial density operator
  of the entangled resource, i.e.,
  \begin{equation}
    \label{eq:commute}
    [\rho_B, U_j] = 0 \quad \mbox{for all $j$ and}
  \end{equation}
\item[(ii)] Alice's measurement eigenstates are obtained from
  $\ket\phi_A$ by the same unitaries as Bob's, more precisely,
  \begin{equation}
    \label{eq:phij}
    \ket{\phi_j}_A = J^\dag U_j^\dag J \ket\phi_A.
  \end{equation}
\end{enumerate}
where $J$ is the antiunitary isomorphism between $\HH_A$ and $\HH_B$
introduced in (\ref{eq:antipolar}).  Indeed, if Alice's measurement
results the state (\ref{eq:phij}) then Bob's state becomes
\begin{eqnarray}
  \ket{\psi_j}_B &=& \frac1{\sqrt p} R \ket{\phi_j}_A
  = \frac1{\sqrt p} \big(\sqrt{\rho_B} J\big)\, 
  \big(J^\dag U_j^\dag J \ket{\phi}_A\big)
  \nonumber\\&=&\label{eq:psii}
  U_j^\dag \left( \frac1{\sqrt p} R \ket{\phi}_A \right)
  = U_j^\dag \ket{\psi}_B
\end{eqnarray}
where the probability $p$ must be the same for every measurement
outcome (i.e., $p=1/N$).  After he received the classical information
about the result, Bob performs the unitary transformation $U_j$ that
just turns (\ref{eq:psii}) to the correct target state $\ket\psi_B$.

We give an example based on Refs.\ \cite{pra65e022316, pra69e022310,
  job7p135} that satisfies the sufficient RSP conditions
(\ref{eq:commute}) and (\ref{eq:phij}).  Let us suppose that Bob's set
of unitary operations are given by
\begin{equation}
  \label{eq:subweyl}
  U_j\ket k = e^{2\pi ijk/N} \ket k
\end{equation}
with $j,k=0,1, \ldots,N-1$.  They form the cyclic group $\mathbb Z_N$
that is a subgroup of the Weyl group used for teleportation.
Condition (\ref{eq:commute}) is immediately fulfilled if the
eigenstates $\ket k_B$ of the unitaries are chosen to be Schmidt
vectors of the entangled state (\ref{eq:Psi}).  It is easy to see that
the measurement eigenstates (\ref{eq:phij}) constitute an orthonormal
basis if and only if $\ket\phi_A$ is an equal-weighted superposition
of all the Schmidt vectors $\ket k_A$:
\begin{equation}
  \label{eq:E_A}
  \ket\phi_A = \frac1{\sqrt N} \sum_{k=0}^{N-1} e^{-i\varphi_k} \ket k_A
\end{equation}
where $\varphi_k$ are the free RSP parameters that are known to Alice
but unknown to Bob and, because of the irrelevant global phase factor,
$N-1$ of them are independent.  Substituting it into \eref{eq:psib}, we
obtain the ensemble of preparable target states
\begin{equation}
  \label{eq:theensemble}
  \mathcal E = \Bigg\{ \ket{\psi} = \sum_{k=0}^{N-1} \alpha_k
  e^{i\varphi_k} \ket{k} \,\Bigg|\, \varphi_k\in\mathbb R\Bigg\}
\end{equation}
which is the $N$ dimensional generalization of the equatorial ensemble
of Ref.~\cite{pra63e014302}.  All the measurement outcomes are equally
probable, that is, the probability distribution is uniform, and Bob
gains no information about the target state from the classical
message.

For a specific element $\ket\psi_B$ in (\ref{eq:theensemble}), the
corresponding measurement basis writes
\begin{equation}
  \label{eq:measbasis}
  \ket{\phi_j}_A = \frac1{\sqrt N} \sum_{k=0}^{N-1} 
  e^{2\pi ijk/N-i\varphi_k} \ket k_A.
\end{equation}
So Alice uses the parameters $\varphi_k$ to tune her measuring
apparatus according to different target states.  A practical way for
her to do it is to perform the measurement always in the same fixed
basis $\ket{\phi_j^{(0)}}_A$ after she applied the unitary
\emph{pre-measurement} transformation $V$ that explicitly depends on
the parameters $\varphi_k$ and generates the basis
(\ref{eq:measbasis}) appropriately.  With our special setup it is
convenient to choose
\begin{eqnarray}
  \ket{\phi_j}_A &=& V^\dag \ket{\phi_j^{(0)}}_A,\\
  \label{eq:jdef}
  \ket{\phi_j^{(0)}}_A &=& \frac1{\sqrt N} \sum_{k=0}^{N-1} 
  e^{2\pi ijk/N} \ket k_A,\\
  \label{eq:V0}
  V &=& \sum_{k=0}^{N-1} e^{i\varphi_k} \ket k_{AA}\bra k.
\end{eqnarray}
When constructing an RSP setup, the goal is to find an implementation
of the pre-measurement transformation and to perform the fixed
measurement that allow deterministic exact RSP of the ensemble
(\ref{eq:theensemble}).

\section{Continuous variable remote state preparation}
\label{sec:cvrsp}

In this section, we extend the idea of \sref{sec:finitersp} to
continuous variables and present example schemes for CVRSP.  We also
investigate whether the proposed schemes are practical.

We have considered RSP schemes having uniform probability distribution
of the measurement outcome.  In the infinite dimensional case with
well-normed (physical) entangled state and discrete variable
measurement (i.e., one having countable infinitely many outcomes, like
photon counting in a mode of the EM field), a discrete probability
distribution cannot be uniform because probabilities should sum up
to~1.  To find a scheme with uniform probability distribution, either
the requirement for a well-normed physical entangled state should be
removed or continuous variable measurement should be allowed, or both.
We shell show examples for all the three cases.

\subsection{RSP by quadrature measurement}
\label{sec:quadrsp}

Let us consider an ideally correlated EPR pair~\cite{pr47p777}, the
same as in ideal CV teleportation.  The wave function of this
unphysical state is
\begin{equation}
  \label{eq:Psiepr}
  \Psi(x_1,x_2) = \frac1{2\pi} \int_{-\infty}^\infty e^{i (x_1-x_2)p}\,dp
  = \delta (x_1-x_2).
\end{equation}
As a generalization of minimum RSP to systems of continuous variables,
one may suppose that Alice measures the momentum of her particle.
Since the total momentum is zero, a measurement result of $p$ would
imply that Bob's particle is in a momentum eigenstate with momentum
$-p$.  Therefore, if Alice wanted to remotely prepare the momentum
eigenstate $\ket{p_0}_B$ at Bob's side, she just has to message Bob to
apply a phase space displacement $\hat D(\alpha)$ on his particle with
$\alpha = 0 + i(p_0+p)$.  The message consists of one real number
($p_0+p$) which yields no information about the target state
$\ket{p_0}_B$ as the probability distribution of the outcome $p$ is
uniform.  The set of unitaries Bob may need to apply is a subgroup of
the Heisenberg--Weyl group of continuous teleportation and is
isomorphic to the additive group of real numbers $\mathbb R$.  Noise
and detector inefficiencies in this scheme can be taken into account
according to Ref.~\cite{job5pS360}.  However, the ensemble of states
that can be prepared this way is rather restricted.  The only
parameter Alice can control is the real number $p_0$ and the protocol
would draw the same results with a classically correlated shared
resource.  It is merely a CV reformulation of the Vernam
cipher (one-time pad) that is a classical cryptographic
protocol using shared randomness as common secret key.

In analogy with \sref{sec:finitersp} and \eref{eq:V0}, we may let
Alice apply a pre-measurement transformation
\begin{equation}
  \label{eq:V1}
  \hat V = e^{i\varphi(\hat X_1)},
\end{equation}
where now the real function $\varphi\colon \mathbb R \to \mathbb R$
plays the role of RSP parameters that are under Alice's control but
are unknown to Bob.  This transforms the entangled state into
\begin{equation}
  \label{eq:Psiepr2}
  (\hat V\Psi)(x_1,x_2) 
  = e^{i\varphi(x_1)} \delta (x_1-x_2).
\end{equation}
Then Alice performs a momentum measurement.  The eigenstate of the
momentum operator $\hat P_1=-i\partial_{x_1}$ corresponding to the
eigenvalue $p$ is given by the wave function $\phi_p^{(0)}(x_1) =
e^{ipx_1}$.  If the outcome $p$ occurs, Bob's conditional state will
be given by the partial inner product $\ket\psi_B = {}_A
\bra{\phi_p^{(0)}}\hat V\ket\Psi_{AB}$
\begin{equation}
  \psi_p(x_2) =  \int_{-\infty}^\infty 
  \phi_p^*(x_1) \Psi(x_1, x_2) \,dx_1
  = e^{i[\varphi(x_2)-px_2]}.
\end{equation}
Bob is provided by the momentum displacement operators $\hat D_2(ip) =
\exp(ip\hat X_2)$ with the real parameter $p$ obtained from the
classical message.  Applying the corresponding operation, he obtains
the correct target state
\begin{equation}
  \label{eq:psix2}
  \psi(x_2) = e^{i\varphi(x_2)}
\end{equation}
that no longer depends on the measurement outcome.  We conclude that
the ensemble of target states that can be prepared using the present
method is
\begin{equation}
  \label{eq:ensemble-quad}
  \mathcal E = \Big\{ \psi(x) = e^{i\varphi(x)}
    \,\Big|\, \varphi\colon \mathbb R \to \mathbb R \Big\}.
\end{equation}

Note that the role of position and momentum can be interchanged, and
any quadrature operators $\hat Q_\theta$ and $\hat Q_{\theta+\pi/4}$
can be used instead of them. 

We also remark that the continuity of the target wave function
(\ref{eq:psix2}) should be ensured by the process that realizes the
pre-measurement transformation (\ref{eq:V1}).  An example in quantum
optics for a possible process is the phase displacement operation
$\hat D(\alpha)$ itself.  It can be realized in a homodyne
interferometric setup by mixing the input mode with an intense
coherent laser beam on a low reflectivity beam splitter.  Since the
displacement parameter $\alpha$ can be tuned by adjusting the phase
and amplitude of the laser field,  one can realize the momentum
displacement operator $\hat D(ip) = \exp (ip\hat X)$ corresponding to
a linear $\varphi(x) = px$.  However, this process alone would lead to
the CV Vernam cipher.

More general pre-measurement transformations are provided by CV
quantum computation.  It can be shown~\cite{prl88e097904} that the
transformation $\hat V = \exp i(\alpha \hat X + \beta \hat X^2)$,
whose generator is a quadratic polynomial in $\hat X$, can be
implemented using appropriate combination of phase space displacements
and squeezing.  Moreover, the cubic phase gate $V = \exp i\gamma \hat
X^3$ can be realized by combining linear optics with the nonlinear
photon number measurement process \cite{pra64e012310, pra65e042304}.
As long as $\hat V$ commutes with $\hat X$, more sophisticated 
\cite{prl82p1784} pre-measurement transformations give rise to larger 
ensemble of preparable states.

To summarize, the proposed CVRSP scheme involves momentum (quadrature)
measurement at Alice's side and phase space displacement operations at
Bob's side.  Remote preparation of the ensemble
(\ref{eq:ensemble-quad}) is possible using the original EPR correlated
resource and the communication of one real number.

\subsection{RSP by phase measurement}
\label{sec:phasersp}

This example is motivated by the fact that Alice's measurement
eigenstates given in \eref{eq:jdef} are analogous to the Pegg--Barnett
phase states in an $N$ dimensional truncated Hilbert
space~\cite{pra39p1665}.  Consider two modes of the EM field.  We
identify states $\ket k_A$ and $\ket k_B$ defined in
\sref{sec:finitersp} with photon number states $\ket n_A$ and $\ket
n_B$ of the corresponding field modes and suppose that Bob can apply
simple optical phase shifts
\begin{equation}
  \label{eq:bobsphaseshift}
  \hat U_\vartheta  = \sum_{n=0}^\infty e^{i\vartheta n} 
  \ket n_{BB} \bra n \equiv e^{i\vartheta \hat N}
\end{equation}
on his mode with phase angles $\vartheta=2\pi j/N$ (for $j=0$,
\ldots,~$N-1$).  We can let Bob take the limit $N\to\infty$
independently of Alice and suppose that he can shift the phase by an
arbitrary amount.  Thus, in the currently proposed CVRSP setup, the
set of Bob's recovering unitary operations is parameterized by a
continuous phase variable, that is, it is isomorphic to the group
$SU(1)$.

According to \eref{eq:commute}, we must choose the entangled shared
resource so that its Schmidt vectors are photon number states.
Although any such resource would do, it is straightforward to consider
a two-mode squeezed vacuum, expressed in the discrete photon number
basis as
\begin{equation}
  \label{eq:2msqvac}
  \ket{\text{sq}}_{AB} = \frac1{\cosh r} \sum_{n=0}^\infty (\tanh r)^n
  \ket n_A \ket n_B.
\end{equation}

Also, we have that Alice's pre-measuring unitary operation
is a phase shift depending on the photon number
\begin{equation}
  \label{eq:Vkerr}
  \hat V = \sum_{n=0}^\infty e^{i\varphi_n} \ket n_{AA} \bra n 
  \equiv e^{i\varphi(\hat N)},
\end{equation}
where $\varphi_n$ or $\varphi(n)$ now denotes a series of real
numbers.  Theoretically, the angles $\varphi_n$ of the phase shift can
be arbitrary.  They are the free RSP parameters that are chosen by
Alice but Bob is unaware of them.  In practice, however, their choice
is rather restricted.  We may consider an optical phase shifter as in
\eref{eq:bobsphaseshift} for which $\varphi_n = \vartheta n$ is linear
in $n$, or a Kerr nonlinear medium \cite{prl82p1784} which
typically realizes a shift $\varphi_n=\chi n^2$ of the quantum
mechanical phase that is quadratic in the photon number $n$.  We can
also combine different methods to achieve more complex functions like
$\varphi_n = \chi n^2 + \vartheta n$.

Let Alice measure the $N$ dimensional truncated Pegg--Barnett
phase states \cite{pra39p1665} in analogy with \eref{eq:jdef}
\begin{equation}
  \label{eq:phasestate}
  \ket{\vartheta_j}_A = \frac1{\sqrt N} \sum_{n=0}^{N-1} 
  e^{2\pi ijn/N} \ket n_A.
\end{equation}
If Alice communicates the result of her measurement, i.e., the angle
$\vartheta_j =2\pi j/N$ to Bob and he applies the corresponding phase
shift~(\ref{eq:bobsphaseshift}), the state prepared becomes
\begin{equation}
  \label{eq:psi0Nfinal}
  \ket{\psi}_B = \frac1{\cosh r\sqrt{p N}} \sum_{n=0}^{N-1}
  (\tanh r)^n e^{i\varphi_n} \ket n_B.
\end{equation}

Finally, in accordance with the Pegg--Barnett principle, we take the limit
$N\to\infty$ at the very end of the procedure, after the output state
was calculated.  The ensemble of states that can be prepared this way
is
\begin{equation}
  \label{eq:ensemble-phase}
  \mathcal E = \Bigg\{ \ket{\psi} 
  =  \sum_{n=0}^\infty \frac{(\tanh r)^n}{\cosh r}
  e^{i\varphi_n}  \ket n
  \,\Bigg|\, \varphi_n \in \mathbb R \Bigg\}.
\end{equation}
The squeezing parameter $r$ is fixed in the protocol, while the real
numbers $\varphi_n$ are under Alice's control and unknown to Bob.  The
transmitted classical information consists of one real number, the
angle $\vartheta$.

Note that the measurement of the truncated Pegg--Barnett phase states
(\ref{eq:phasestate}) does not always yield a truncated phase eigenstate.
It is because they form a complete basis only in the truncated Hilbert
space, and the shared resource~(\ref{eq:2msqvac}) has components of
photon numbers in mode A higher than $N-1$.  Therefore, some of the
possible outcomes of an experiment should be discarded.  The total
probability of success is
\begin{equation}
  \label{eq:Psucc}
  P= N p = \sum_{n=0}^{N-1} 
  \frac{\tanh^{2n}r}{\cosh^2r}=
  1-\tanh^{2N}r
\end{equation}
that tends to 1 as $N$ increases, that is, the probability of failure
decreases exponentially with $N$.  

Recently, schemes have been developed to experimentally perform the
direct single-shot measurement of truncated phase eigenstates in a
probabilistic way using beam splitters, mirrors, phase shifters and
photodetectors \cite{prl76p4148, prl89e173601, pra67p063814,
  pla323p329}.  These setups involve $N$ auxiliary modes and the
number of the optical elements required scales polynomially with $N$.
Although, they realize POVM measurements, the state projections of our
needs can be obtained.  However, the probability of success decreases
with $N$.  Therefore, at present state of art, an experimental
realization of our CVRSP scheme may be applicable for preparation of
states not containing too large photon numbers.

To summarize, the proposed CVRSP scheme utilizes two-mode squeezed
vacuum as shared entangled resource.  Alice performs a unitary
pre-measurement operation shifting the phase by an amount depending on
the photon number.  This can be (but not restricted to), e.g., either
or both (i) a constant phase shift $\vartheta$ on each photon or (ii)
a Kerr-like nonlinear phase shift $\varphi_n = \chi n^2$.  Then she
performs a phase measurement, she communicates the result to Bob, and
he applies a linear phase shift on his mode accordingly.  The ensemble
of states that can be prepared this way is given by
\eref{eq:ensemble-phase}.

\subsection{RSP by photon counting}
\label{sec:photonrsp}

Now we present a scheme that involves discrete photon number
measurement.  According to \sref{sec:cvrsp}, we release the
requirement for a well-normed entangled state and consider the
following unphysical state
\begin{equation}
  \ket\Psi_{AB} = \sum_{n=0}^\infty \ket n_A \ket n_B.
\end{equation}
As the first step, suppose that Alice wants to remotely prepare the
photon number state $\ket m_B$.  For this purpose, she applies the
non-unitary down-shift operator
\begin{equation}
  \label{eq:downshift}
  \hat U_m = \sum_{n=0}^\infty \ket n\bra{n+m}
  \equiv [(\hat n+1)^{-1/2} \hat a]^m 
  \equiv e^{im\hat\Phi}
\end{equation}
on system A that decreases the photon number by $m$ dropping out all
components with number of photons less than $m$.  Note that $\hat U$
is a simple exponential function of the non-Hermitian phase operator
$\hat\Phi$.  This way, Alice turns the shared state into
\begin{equation}
  \ket{\Psi'}_{AB} = \sum_{n=m}^\infty \ket {n-m}_A \ket n_B
  = \sum_{n=0}^\infty \ket n_A \ket {n+m}_B.
\end{equation}
Now she performs a photon counting measurement and communicates the
result $n$ to Bob.  Since his conditional state is $\ket{n+m}_B$, he
just has to apply the down-shift operation $\hat U_n$ to obtain the
correct target state $\ket m_B$.

Though the above example is a simple infinite dimensional extension of
the classical Vernam cipher, we will show in a rigorous way that, by
letting Alice apply more complex pre-measurement operations instead of
(\ref{eq:downshift}), we can enlarge the ensemble of preparable states
and turn the scheme into a CVRSP-like protocol.

Let us consider first the $N$ dimensional truncated entangled resource
\begin{equation}
  \label{eq:Psiphase}
  \ket{\Psi}_{AB} = \frac1{\sqrt N} \sum_{j=0}^{N-1}
  \ket{\vartheta_j}_A  \ket{-\vartheta_j}_B
  = \frac1{\sqrt N} \sum_{n=1}^{N-1}
  \ket{n}_A   \ket{n}_B,
\end{equation}
where we used the shorthand $-\vartheta_j$ for $\vartheta_{N-j} =
2\pi-\vartheta_j$.  The total phase of the two modes and the relative
photon number are both zero.  Now we chose Bob's unitaries to have the
phase states $\ket{\vartheta_j}_B$ as eigenstates
\begin{equation}
  \label{eq:bobsladder}
  \begin{aligned}
    U_n \ket{\vartheta_j}_B 
    &= e^{2\pi ijn/N} \ket{\vartheta_j}_B,
    \\ 
    U_n \ket{k}_B &= \ket{k\ominus n}_B,
  \end{aligned}
\end{equation}
If $\Phi$ denotes the Hermitian phase operator in the
$N$ dimensional truncated Hilbert space then $U_1$ corresponds to
$\exp(i\Phi)$, and $U_n = \exp (i n \Phi)$ and there is no problem
with the unitarity of $U_n$.

We define Alice's pre-measuring unitary operation as
\begin{equation}
  \label{eq:V2}
  V = \sum_{j=0}^{N-1} e^{i\varphi_j} 
  \ket {\vartheta_j}_{AA} \bra {\vartheta_j}
  \equiv e^{i\varphi(\Phi)}
  = \sum_{n=0}^{N-1} f_n U_n,
\end{equation}
where the coefficients $f_n$ are the well-normed discrete Fourier
series of the phase shifts $e^{i\varphi_j}$
\begin{equation}
  \label{eq:fn-series}
  f_n = \frac1N \sum_{j=0}^{N-1} e^{-2\pi ijn/N} e^{i\varphi_j},
  \qquad \sum_{n=0}^{N-1} |f_n|^2 = 1.
\end{equation}
Let Alice perform a photon counting measurement with eigenstates
$\ket n_A$.  After the measurement and the corresponding unitary
transformation $U_n$ we obtain that the following state is
prepared remotely at Bob's site
\begin{equation}
  \label{eq:psiNfinal2}
  \ket{\psi}_B = \frac1{\sqrt N} \sum_{j=0}^{N-1}
  e^{i\varphi_j} \ket {-\vartheta_j}_B 
  = \sum_{n=0}^{N-1} f_n \ket n_B.
\end{equation}

In the limit $N\to\infty$, we can use the unnormed
continuous phase eigenbasis
\begin{equation}
  \ket\vartheta = \sum_{n=0}^\infty e^{i\vartheta n} \ket n
\end{equation}
to express the unphysical maximally entangled EPR state
\begin{equation}
  \label{eq:phaseEPR}
  \ket\Psi_{AB} = \frac1{2\pi} \int_0^{2\pi}
  \ket\vartheta_A \ket{-\vartheta}_B \,d\vartheta
  = \sum_{n=0}^\infty \ket n_A \ket n_B.
\end{equation}
$U_1$ turns into the non-unitary $\hat U_1 = \exp(i\hat\Phi) = (\hat
n+1)^{-1/2} \hat a$ photon number down-shift operator, and
the Fourier series (\ref{eq:fn-series}) turn into
the discrete Fourier transform of the periodic phase shift function
$e^{i\varphi(\vartheta)}$
\begin{equation}
  \label{eq:fn-trans}
  f_n = \frac1{2\pi} \int_0^{2\pi} 
  e^{-in\vartheta} e^{i\varphi(\vartheta)} \,d\vartheta,
  \qquad \sum_{n=0}^\infty |f_n|^2 = 1.
\end{equation}
Alice's non-unitary pre-measuring operator becomes
\begin{equation}
  \label{eq:V3}
  \begin{aligned}
    \hat V &= \int_0^{2\pi} e^{i\varphi(\vartheta)} 
    \ket {\vartheta}_{AA} \bra{\vartheta} \,d\vartheta
    \equiv e^{i\varphi(\hat\Phi)}
    \\&= f_0 \hat\id + \sum_{n=1}^\infty \left( 
      f_n \hat U_n + f_{-n} \hat U_n^\dag \right).
  \end{aligned}
\end{equation}
After Alice applied this pre-measurement transformation, she performs
a photon counting measurement.  If she finds $n$ photons
in mode A, then Bob's conditional state becomes
\begin{equation}
  \ket{\psi_n}_B = \sum_{k=0}^\infty f_{(k-n)} \ket{k}_B
  = \sum_{k=-n}^\infty f_k \ket{k+n}_B.
\end{equation}
Applying the down-shift operator $\hat U_n$, we obtain the target
state that has now become unconditional on the measurement outcome:
\begin{equation}
  \ket\psi_B = \frac1{2\pi} \int_0^{2\pi} e^{i\varphi(\vartheta)}
  \ket{-\vartheta}_B \,d\vartheta
  = \sum_{n=0}^\infty f_n \ket{n}_B,
\end{equation}
where the periodic real function $\varphi(\vartheta)$ or the Fourier
coefficients $f_n$ that are given by (\ref{eq:fn-trans}) serve as the
RSP parameters.

We note that the presented scheme seems to be irrealistic for that it
utilizes unphysical shared state and the photon number down-shift
operation is hard to realize.  The setup is still interesting
theoretically because it shows that unitary operations at Bob's site
are not necessary, irreversible nonunitary operations can also be used
to restore the output state.

\section{Conclusions}
\label{sec:conc}

We have considered exact deterministic remote state preparation with
minimal classical communication.  For finite $N$ dimensional systems, 
this corresponds to a projective measurement at Alice's side, 
communication of the minimal amount of $\log N$ bits of classical 
information, and a unitary transformation at Bob's side.  The ensemble 
of preparable states is parameterized by $N-1$ real numbers called
RSP parameters which are under Alice's control and unknown to
Bob.

We have introduced continuous variable versions of the above remote
state preparation analogous to continuous variable teleportation.
While in the latter, two real numbers are messaged, the classical
communication cost of our CVRSP schemes is one real number only.
Still, there are infinitely many RSP parameters that determine the
target state.  We have proposed three example setups consisting of (i)
momentum measurement and momentum displacement operation, (ii) phase
measurement and phase shifters, and (iii) photon number measurement
and photon number shift operations.  Alice's unitary pre-measurement
transformation, into which the RSP parameters are fed, is function of
the variable canonically conjugated to what is measured, i.e., it is
function of (i) the position operator, (ii) the number operator, and
(iii) the non-Hermitian phase operator, respectively.


\begin{acknowledgments}
  Our research was partially supported by the National Research Fund
  of Hungary (OTKA) under Contract Nos.\ T043287 and T049234 and by
  the Hungarian Ministry of Education under contract No. CZ-5/03
  (Hungarian-Czech scientific cooperation).
\end{acknowledgments}


\begin{thebibliography}{29}
\expandafter\ifx\csname natexlab\endcsname\relax\def\natexlab#1{#1}\fi
\expandafter\ifx\csname bibnamefont\endcsname\relax
  \def\bibnamefont#1{#1}\fi
\expandafter\ifx\csname bibfnamefont\endcsname\relax
  \def\bibfnamefont#1{#1}\fi
\expandafter\ifx\csname citenamefont\endcsname\relax
  \def\citenamefont#1{#1}\fi
\expandafter\ifx\csname url\endcsname\relax
  \def\url#1{\texttt{#1}}\fi
\expandafter\ifx\csname urlprefix\endcsname\relax\def\urlprefix{URL }\fi
\providecommand{\bibinfo}[2]{#2}
\providecommand{\eprint}[2][]{\url{#2}}

\bibitem[{\citenamefont{Bennett et~al.}(1993)\citenamefont{Bennett, Brassard,
  Cr\'epeau, Jozsa, Peres, and Wootters}}]{prl70p1895}
\bibinfo{author}{\bibfnamefont{C.~H.} \bibnamefont{Bennett}},
  \bibinfo{author}{\bibfnamefont{G.}~\bibnamefont{Brassard}},
  \bibinfo{author}{\bibfnamefont{C.}~\bibnamefont{Cr\'epeau}},
  \bibinfo{author}{\bibfnamefont{R.}~\bibnamefont{Jozsa}},
  \bibinfo{author}{\bibfnamefont{A.}~\bibnamefont{Peres}}, \bibnamefont{and}
  \bibinfo{author}{\bibfnamefont{W.~K.} \bibnamefont{Wootters}},
  \bibinfo{journal}{Phys. Rev. Lett.} \textbf{\bibinfo{volume}{70}},
  \bibinfo{pages}{1895} (\bibinfo{year}{1993}).

\bibitem[{\citenamefont{Bennett et~al.}(2005)\citenamefont{Bennett, Hayden,
  Leung, Shor, and Winter}}]{tit51p56}
\bibinfo{author}{\bibfnamefont{C.~H.} \bibnamefont{Bennett}},
  \bibinfo{author}{\bibfnamefont{P.}~\bibnamefont{Hayden}},
  \bibinfo{author}{\bibfnamefont{D.~W.} \bibnamefont{Leung}},
  \bibinfo{author}{\bibfnamefont{P.~W.} \bibnamefont{Shor}}, \bibnamefont{and}
  \bibinfo{author}{\bibfnamefont{A.}~\bibnamefont{Winter}},
  \bibinfo{journal}{IEEE Trans. Inform. Theory} \textbf{\bibinfo{volume}{51}},
  \bibinfo{pages}{56} (\bibinfo{year}{2005}), \eprint{quant-ph/0307100}.

\bibitem[{\citenamefont{Lo}(2000)}]{pra62e012313}
\bibinfo{author}{\bibfnamefont{H.~K.} \bibnamefont{Lo}},
  \bibinfo{journal}{Phys. Rev. A} \textbf{\bibinfo{volume}{62}},
  \bibinfo{pages}{012313} (\bibinfo{year}{2000}).

\bibitem[{\citenamefont{Devetak and Berger}(2001)}]{prl87e197901}
\bibinfo{author}{\bibfnamefont{I.}~\bibnamefont{Devetak}} \bibnamefont{and}
  \bibinfo{author}{\bibfnamefont{T.}~\bibnamefont{Berger}},
  \bibinfo{journal}{Phys. Rev. Lett.} \textbf{\bibinfo{volume}{87}},
  \bibinfo{pages}{197901} (\bibinfo{year}{2001}).

\bibitem[{\citenamefont{Leung and Shor}(2003)}]{prl90e127905}
\bibinfo{author}{\bibfnamefont{D.~W.} \bibnamefont{Leung}} \bibnamefont{and}
  \bibinfo{author}{\bibfnamefont{P.~W.} \bibnamefont{Shor}},
  \bibinfo{journal}{Phys. Rev. Lett.} \textbf{\bibinfo{volume}{90}},
  \bibinfo{pages}{127905} (\bibinfo{year}{2003}).

\bibitem[{\citenamefont{Hayashi et~al.}(2003)\citenamefont{Hayashi, Hashimoto,
  and Horibe}}]{pra67e052302}
\bibinfo{author}{\bibfnamefont{A.}~\bibnamefont{Hayashi}},
  \bibinfo{author}{\bibfnamefont{T.}~\bibnamefont{Hashimoto}},
  \bibnamefont{and} \bibinfo{author}{\bibfnamefont{M.}~\bibnamefont{Horibe}},
  \bibinfo{journal}{Phys. Rev. A} \textbf{\bibinfo{volume}{67}},
  \bibinfo{pages}{052302} (\bibinfo{year}{2003}).

\bibitem[{\citenamefont{Abeyesinghe and Hayden}(2003)}]{pra68e062319}
\bibinfo{author}{\bibfnamefont{A.}~\bibnamefont{Abeyesinghe}} \bibnamefont{and}
  \bibinfo{author}{\bibfnamefont{P.}~\bibnamefont{Hayden}},
  \bibinfo{journal}{Phys. Rev. A} \textbf{\bibinfo{volume}{68}},
  \bibinfo{pages}{062319} (\bibinfo{year}{2003}).

\bibitem[{\citenamefont{Pati}(2000)}]{pra63e014302}
\bibinfo{author}{\bibfnamefont{A.~K.} \bibnamefont{Pati}},
  \bibinfo{journal}{Phys. Rev. A} \textbf{\bibinfo{volume}{63}},
  \bibinfo{pages}{014302} (\bibinfo{year}{2000}).

\bibitem[{\citenamefont{Zeng and Zhang}(2002)}]{pra65e022316}
\bibinfo{author}{\bibfnamefont{B.}~\bibnamefont{Zeng}} \bibnamefont{and}
  \bibinfo{author}{\bibfnamefont{P.}~\bibnamefont{Zhang}},
  \bibinfo{journal}{Phys. Rev. A} \textbf{\bibinfo{volume}{65}},
  \bibinfo{pages}{022316} (\bibinfo{year}{2002}).

\bibitem[{\citenamefont{Ye et~al.}(2004)\citenamefont{Ye, Zhang, and
  Guo}}]{pra69e022310}
\bibinfo{author}{\bibfnamefont{M.-Y.} \bibnamefont{Ye}},
  \bibinfo{author}{\bibfnamefont{Y.-S.} \bibnamefont{Zhang}}, \bibnamefont{and}
  \bibinfo{author}{\bibfnamefont{G.-C.} \bibnamefont{Guo}},
  \bibinfo{journal}{Phys. Rev. A} \textbf{\bibinfo{volume}{69}},
  \bibinfo{pages}{022310} (\bibinfo{year}{2004}).

\bibitem[{\citenamefont{Kurucz and Adam}(2005)}]{job7p135}
\bibinfo{author}{\bibfnamefont{Z.}~\bibnamefont{Kurucz}} \bibnamefont{and}
  \bibinfo{author}{\bibfnamefont{P.}~\bibnamefont{Adam}}, \bibinfo{journal}{J.
  Opt. B: Quantum Semiclass. Opt.} \textbf{\bibinfo{volume}{7}},
  \bibinfo{pages}{135} (\bibinfo{year}{2005}).

\bibitem[{\citenamefont{Lloyd and Braunstein}(1999)}]{prl82p1784}
\bibinfo{author}{\bibfnamefont{S.}~\bibnamefont{Lloyd}} \bibnamefont{and}
  \bibinfo{author}{\bibfnamefont{S.~L.} \bibnamefont{Braunstein}},
  \bibinfo{journal}{Phys. Rev. Lett.} \textbf{\bibinfo{volume}{82}},
  \bibinfo{pages}{1784} (\bibinfo{year}{1999}).

\bibitem[{\citenamefont{Gottesman et~al.}(2001)\citenamefont{Gottesman, Kitaev,
  and Preskill}}]{pra64e012310}
\bibinfo{author}{\bibfnamefont{D.}~\bibnamefont{Gottesman}},
  \bibinfo{author}{\bibfnamefont{A.}~\bibnamefont{Kitaev}}, \bibnamefont{and}
  \bibinfo{author}{\bibfnamefont{J.}~\bibnamefont{Preskill}},
  \bibinfo{journal}{Phys. Rev. A} \textbf{\bibinfo{volume}{64}},
  \bibinfo{pages}{012310} (\bibinfo{year}{2001}).

\bibitem[{\citenamefont{Bartlett and Sanders}(2002)}]{pra65e042304}
\bibinfo{author}{\bibfnamefont{S.~D.} \bibnamefont{Bartlett}} \bibnamefont{and}
  \bibinfo{author}{\bibfnamefont{B.~C.} \bibnamefont{Sanders}},
  \bibinfo{journal}{Phys. Rev. A} \textbf{\bibinfo{volume}{65}},
  \bibinfo{pages}{042304} (\bibinfo{year}{2002}).

\bibitem[{\citenamefont{Bartlett et~al.}(2002)\citenamefont{Bartlett, Sanders,
  Braunstein, and Nemoto}}]{prl88e097904}
\bibinfo{author}{\bibfnamefont{S.~D.} \bibnamefont{Bartlett}},
  \bibinfo{author}{\bibfnamefont{B.~C.} \bibnamefont{Sanders}},
  \bibinfo{author}{\bibfnamefont{S.~L.} \bibnamefont{Braunstein}},
  \bibnamefont{and} \bibinfo{author}{\bibfnamefont{K.}~\bibnamefont{Nemoto}},
  \bibinfo{journal}{Phys. Rev. Lett.} \textbf{\bibinfo{volume}{88}},
  \bibinfo{pages}{097904} (\bibinfo{year}{2002}).

\bibitem[{\citenamefont{Bartlett and Munro}(2003)}]{prl90e117901}
\bibinfo{author}{\bibfnamefont{S.~D.} \bibnamefont{Bartlett}} \bibnamefont{and}
  \bibinfo{author}{\bibfnamefont{W.~J.} \bibnamefont{Munro}},
  \bibinfo{journal}{Phys. Rev. Lett.} \textbf{\bibinfo{volume}{90}},
  \bibinfo{pages}{117901} (\bibinfo{year}{2003}).

\bibitem[{\citenamefont{Vaidman}(1994)}]{pra49p1473}
\bibinfo{author}{\bibfnamefont{L.}~\bibnamefont{Vaidman}},
  \bibinfo{journal}{Phys. Rev. A} \textbf{\bibinfo{volume}{49}},
  \bibinfo{pages}{1473} (\bibinfo{year}{1994}).

\bibitem[{\citenamefont{Braunstein and Kimble}(1998)}]{prl80p869}
\bibinfo{author}{\bibfnamefont{S.~L.} \bibnamefont{Braunstein}}
  \bibnamefont{and} \bibinfo{author}{\bibfnamefont{H.~J.}
  \bibnamefont{Kimble}}, \bibinfo{journal}{Phys. Rev. Lett.}
  \textbf{\bibinfo{volume}{80}}, \bibinfo{pages}{869} (\bibinfo{year}{1998}).

\bibitem[{\citenamefont{Furusawa et~al.}(1998)\citenamefont{Furusawa,
  S\/orensen, Braunstein, Fuchs, Kimble, and Polzik}}]{sci282p706}
\bibinfo{author}{\bibfnamefont{A.}~\bibnamefont{Furusawa}},
  \bibinfo{author}{\bibfnamefont{J.~L.} \bibnamefont{S\/orensen}},
  \bibinfo{author}{\bibfnamefont{S.~L.} \bibnamefont{Braunstein}},
  \bibinfo{author}{\bibfnamefont{C.~A.} \bibnamefont{Fuchs}},
  \bibinfo{author}{\bibfnamefont{H.~J.} \bibnamefont{Kimble}},
  \bibnamefont{and} \bibinfo{author}{\bibfnamefont{E.~S.}
  \bibnamefont{Polzik}}, \bibinfo{journal}{Science}
  \textbf{\bibinfo{volume}{282}}, \bibinfo{pages}{706} (\bibinfo{year}{1998}).

\bibitem[{\citenamefont{Einstein et~al.}(1935)\citenamefont{Einstein, Podolsky,
  and Rosen}}]{pr47p777}
\bibinfo{author}{\bibfnamefont{A.}~\bibnamefont{Einstein}},
  \bibinfo{author}{\bibfnamefont{B.}~\bibnamefont{Podolsky}}, \bibnamefont{and}
  \bibinfo{author}{\bibfnamefont{N.}~\bibnamefont{Rosen}},
  \bibinfo{journal}{Phys. Rev.} \textbf{\bibinfo{volume}{47}},
  \bibinfo{pages}{777} (\bibinfo{year}{1935}).

\bibitem[{\citenamefont{Paris et~al.}(2003)\citenamefont{Paris, Cola, and
  Bonifacio}}]{job5pS360}
\bibinfo{author}{\bibfnamefont{M.~G.~A.} \bibnamefont{Paris}},
  \bibinfo{author}{\bibfnamefont{M.}~\bibnamefont{Cola}}, \bibnamefont{and}
  \bibinfo{author}{\bibfnamefont{R.}~\bibnamefont{Bonifacio}},
  \bibinfo{journal}{J. Opt. B: Quantum Semiclass. Opt.}
  \textbf{\bibinfo{volume}{5}}, \bibinfo{pages}{S360} (\bibinfo{year}{2003}).

\bibitem[{\citenamefont{Uhlmann}(2000)}]{lnp539p93}
\bibinfo{author}{\bibfnamefont{A.}~\bibnamefont{Uhlmann}}, in
  \emph{\bibinfo{booktitle}{Lecture Notes on Physics}}, edited by
  \bibinfo{editor}{\bibfnamefont{A.}~\bibnamefont{Borowiec}}
  (\bibinfo{publisher}{Springer-Verlag}, \bibinfo{address}{Berlin},
  \bibinfo{year}{2000}), vol. \bibinfo{volume}{539}, pp.
  \bibinfo{pages}{93--105}, \eprint{quant-ph/9901027}.

\bibitem[{\citenamefont{Kurucz et~al.}(2001)\citenamefont{Kurucz, Koniorczyk,
  and Janszky}}]{fdp49p1019}
\bibinfo{author}{\bibfnamefont{Z.}~\bibnamefont{Kurucz}},
  \bibinfo{author}{\bibfnamefont{M.}~\bibnamefont{Koniorczyk}},
  \bibnamefont{and} \bibinfo{author}{\bibfnamefont{J.}~\bibnamefont{Janszky}},
  \bibinfo{journal}{Fortschr. Phys.} \textbf{\bibinfo{volume}{49}},
  \bibinfo{pages}{1019} (\bibinfo{year}{2001}), \eprint{quant-ph/0308020}.

\bibitem[{\citenamefont{Kurucz et~al.}(2003)\citenamefont{Kurucz, Koniorczyk,
  Adam, and Janszky}}]{job5pS627}
\bibinfo{author}{\bibfnamefont{Z.}~\bibnamefont{Kurucz}},
  \bibinfo{author}{\bibfnamefont{M.}~\bibnamefont{Koniorczyk}},
  \bibinfo{author}{\bibfnamefont{P.}~\bibnamefont{Adam}}, \bibnamefont{and}
  \bibinfo{author}{\bibfnamefont{J.}~\bibnamefont{Janszky}},
  \bibinfo{journal}{J. Opt. B: Quantum Semiclass. Opt.}
  \textbf{\bibinfo{volume}{5}}, \bibinfo{pages}{S627} (\bibinfo{year}{2003}).

\bibitem[{\citenamefont{Pegg and Barnett}(1989)}]{pra39p1665}
\bibinfo{author}{\bibfnamefont{D.~T.} \bibnamefont{Pegg}} \bibnamefont{and}
  \bibinfo{author}{\bibfnamefont{S.~M.} \bibnamefont{Barnett}},
  \bibinfo{journal}{Phys. Rev. A} \textbf{\bibinfo{volume}{39}},
  \bibinfo{pages}{1665} (\bibinfo{year}{1989}).

\bibitem[{\citenamefont{Barnett and Pegg}(1996)}]{prl76p4148}
\bibinfo{author}{\bibfnamefont{S.~M.} \bibnamefont{Barnett}} \bibnamefont{and}
  \bibinfo{author}{\bibfnamefont{D.~T.} \bibnamefont{Pegg}},
  \bibinfo{journal}{Phys. Rev. Lett.} \textbf{\bibinfo{volume}{76}},
  \bibinfo{pages}{4148} (\bibinfo{year}{1996}).

\bibitem[{\citenamefont{Pregnell and Pegg}(2002)}]{prl89e173601}
\bibinfo{author}{\bibfnamefont{K.~L.} \bibnamefont{Pregnell}} \bibnamefont{and}
  \bibinfo{author}{\bibfnamefont{D.~T.} \bibnamefont{Pegg}},
  \bibinfo{journal}{Phys. Rev. Lett.} \textbf{\bibinfo{volume}{89}},
  \bibinfo{pages}{173601} (\bibinfo{year}{2002}).

\bibitem[{\citenamefont{Pregnell and Pegg}(2003)}]{pra67p063814}
\bibinfo{author}{\bibfnamefont{K.~L.} \bibnamefont{Pregnell}} \bibnamefont{and}
  \bibinfo{author}{\bibfnamefont{D.~T.} \bibnamefont{Pegg}},
  \bibinfo{journal}{Phys. Rev. A} \textbf{\bibinfo{volume}{67}},
  \bibinfo{pages}{063814} (\bibinfo{year}{2003}).

\bibitem[{\citenamefont{Zou et~al.}(2004)\citenamefont{Zou, Pahlke, and
  Mathis}}]{pla323p329}
\bibinfo{author}{\bibfnamefont{X.}~\bibnamefont{Zou}},
  \bibinfo{author}{\bibfnamefont{K.}~\bibnamefont{Pahlke}}, \bibnamefont{and}
  \bibinfo{author}{\bibfnamefont{W.}~\bibnamefont{Mathis}},
  \bibinfo{journal}{Phys. Lett. A} \textbf{\bibinfo{volume}{323}},
  \bibinfo{pages}{329} (\bibinfo{year}{2004}).

\end{thebibliography}
\end{document}